\documentclass[useAMS,usenatbib]{mn2e}
\usepackage{natbib}
\usepackage{graphicx}
\usepackage{savesym}
\usepackage{listings}
\usepackage[intlimits]{amsmath}
\savesymbol{iint}
\usepackage{txfonts}
\usepackage{bm}
\restoresymbol{TXF}{iint}
\usepackage{amssymb}
\usepackage{hyperref}
\usepackage{amsmath}
\usepackage{tabularx} 
\usepackage{color}
\usepackage{soul}
\usepackage{ulem}

\def\apjl{Astrophys. J. Lett.}
\def\apjs{Astrophys. J. Supp. Ser. }

\def\aap{Astron. Astrophys. }

\newcommand{\vv}[1]{{\bf #1}}
\newcommand{\VV}[1]{\overrightarrow{#1}}

\newcommand{\beq}{\begin{equation}}
\newcommand{\eeq}{\end{equation}}

\newcommand{\spr}[2]{{\bmath #1} \!\cdot\! {\bmath #2}}
\newcommand{\vpr}[2]{{\bmath #1} \!\times\! {\bmath #2}}

\newcommand{\Fs}{\,^*\! F}

\newcommand{\cE}{\bmath{\check{E}}}
\newcommand{\cB}{\bmath{\check{B}}}

\newcommand{\bE}{\bmath{E}}

\title[Electrically-charged black holes]{Electrically-charged black holes  and the Blandford-Znajek mechanism}
\author[Serguei S. Komissarov]{Serguei S. Komissarov$^{1}$ \thanks{Email: S.S.Komissarov@leeds.ac.uk}\\
$^{1}$School of Mathematics, University of Leeds, Leeds, LS2 9JT, UK}

\begin{document}

\maketitle
\date{\today}

\begin{abstract}
Recently, it was claimed that accretion of electric charge by a black hole rotating in an aligned external magnetic field results in a ``dead'' vacuum magnetosphere, where the electric field is totally screened, no vacuum breakdown is possible, and the Blandford-Znajek mechanism cannot operate \citep{KP21}. Here we study in details the properties of the \citet{Wald74} solution for electrically charged black holes discussed in that paper.  Our results show that the claim is erroneous as in the solution with the critical charge $q_0=2aB_0$ there exists a drop of electrostatic potential along all magnetic field lines except the one coinciding with the symmetry axis. It is also found that while uncharged rotating black holes expel external vacuum magnetic field from their event horizon (the Meissner effect), electric charging of black holes pulls the magnetic field lines back on it, resembling what has been observed in some previous force-free, RMHD and PIC simulations of black hole magnetospheres. This suggests that accretion of electric charge may indeed be a feature of the black hole electrodynamics.  However, our analysis shows that the value $q_0$ of the BH charge given by Wald is likely to be only an upper limit, and that the actual value depends of the details of the magnetospheric physics.    
\end{abstract}

\section{Introduction}

In 1974, Robert M. Wald published a paper containing an exact axisymmetric solution for a  magnetosphere of rotating black hole \citep{Wald74}. The solution describes asymptotically uniform magnetic field of finite strength $B_0$ which is aligned with the symmetry axis of the black hole (BH).  The most important feature of this solution is the gravitationally induced electric field in the vicinity of the BH. In particular, \citet{Wald74} showed that this field can accelerate charged particles sliding along the magnetic field line coinciding with symmetry axis.  In this regard, the BH magnetosphere is similar to the magnetosphere of a pulsar, where the electric field is induced by the rotation of a neutron star.          
The exceptionally high strength of pulsar magnetic field leads to very high energies of electrically charged particles accelerated by the rotationally induced electric field and allows vacuum breakdown via copious pair production. As the result, pulsar  magnetosphere is filled with pair plasma that is magnetically driven away in the form of the pulsar wind. The wind extracts the rotational energy of the neutron star mostly in the form of  Poynting flux. \citet{BZ-77} drew analogy between BHs and pulsars and looked for ways to show that the rotational energy of BHs can also be extracted electromagnetically. In particular, they succeeded in finding an analytic solution describing an electromagnetic wind from a Kerr black hole with magnetic charge.  For this, they used the force-free approximation of magnetically-dominated plasma where the particle inertia is negligibly small and but the electric conductivity is high (perfect).  Since than, the Blandford-Znajek (BZ) mechanism has been accepted as one of the main channels for powering activity of galactic nuclei (AGN) and other astrophysical phenomena associated with astrophysical BHs.   In particular, it is believed that this mechanism is behind the acceleration of relativistic jets from AGN.  

One of the main aims of \citet{Wald74} was the question of how  BHs may acquire net electric charge. This is an important issue because the electric charge is the only way for BHs to develop their own electromagnetic field.  By analysing his solution along the symmetry axis, Wald showed that, along the magnetic field line coinciding with the axis, the gravitationally induced electric field pulled charged particles of one particular sign into the BH  and ejected particles of the opposite sign. Citing the results by \citet{Carter73} (republished as \citet{Carter10}), Wald claimed that the event horizon was an equipotential surface and hence the conclusion applied to any magnetic field line crossing the horizon.  
By analysing the modified solution where the BH is allowed to have an electric charge, \citet{Wald74} concluded that the charge ``accretion'' continues until the net charge of the BH reaches some critical value $q_0$. At this point, the horizon potential is the same as at infinity.  As far as the BZ mechanism is concerned, this conclusion by Wald means that the BH magnetosphere dies -- it is no longer able to accelerate charged particles and drive the vacuum breakdown.  Even if such magnetosphere is filled with $e^-\!-\!e^+$ pairs via the two-photon process involving high-energy photons emitted by the accretion disk and its corona, it remains dead because poloidal electric field is needed to drive the poloidal electric currents of the BZ-mechanism. 

Recently, this conclusion was emphasised by \citet{KP21}, who also pointed that once the critical charge is reached, the electromagnetic potential is given by the same 4-vector equation as in the Wald solution for uncharged non-rotating BH.  Since the BZ mechanism does not operate for such BHs, \citet{KP21} argued that it cannot operate for rotating BHs with the critical electric charge too. Hence they argued that AGN jets must be driven by magnetised accretion disks. 

The conclusions by \citet{Wald74} and \citet{KP21} is in conflict with numerous MHD and force-free simulations of black hole magnetospheres, as well as with the few recent PIC simulations \citep{Parfrey19,Crinquand20}. It is also in conflict with the theoretical analysis by \citet{ssk-ebh} which shows that the gravitationally induced electric field of any steady-state axisymmetric BH magnetosphere cannot be screened provided the magnetosphere has no poloidal electric current.  

These conclusions also contradict to the membrane paradigm of BHs, which states that their event horizon is analogous to a rotating conducting sphere \citep{TPM86}. The surface of a conductor is indeed equipotential when not threaded by magnetic field. If, however, it is rotating and threaded by magnetic field, the $q_p\vpr{v}{B}$ force causes the electric charge separation that results in variation of the electrostatic potential over the surface. The classic example of this process is the Faraday disk.  This conclusion applies to conductors with non-vanishing net electric charge too.  

Given the importance of the Blandford-Znajek mechanism in astrophysics of black holes, the conflict needs to be resolved and this is the main objective of our investigation described in this paper. In Section \ref{sec:WS}, we describe the Wald solution for electrically charged rotating BHs   and explain the arguments of \citet{Wald74} and \citet{KP21} in details. 
The analysis of this solution, which exposes the flaws of these arguments and shows that the gravitationally induced electric field does not become screened in BHs with the critical electric charge  is given in  Section  \ref{sec:PWS}.  Section \ref{sec:GC}    
describes how this conclusion is geralised to the general case. Section \ref{sec:Disc} is a discussion of these results and their implications.

In the equations, we utilise the relativistic units where the speed of light $c=1$, the black hole mass $m=1$, and $4\pi$ does not appear in Maxwell's equations.

\section{Wald solutions}
\label{sec:WS}

Wald vacuum solution for the electromagnetic 4-potential is

\beq
   U_\mu = \frac{B_0}{2}(m_\mu+2a k_\mu) \,,
   \label{eq:wald-kerr}
\eeq
where $\VV{m}=\partial_\phi$ ($m^\mu=\delta^\mu_\phi$) and $\VV{k}=\partial_t$ ($k^\mu=\delta^\mu_t$) are the Killing vectors of the stationary axisymmetric spacetime of BH. The electric charge of the BH vanishes, and at infinity the magnetic field is uniform, has the strength of  $B_0$, and is aligned with the BH symmetry axis.  

For a non-rotating BH ($a=0$), equation \eqref{eq:wald-kerr} reduces to 
\beq
   U_\mu = \frac{B_0}{2}m_\mu \,.
   \label{eq:wald-schw}
\eeq

Wald noticed that the solution \eqref{eq:wald-kerr} implies a drop of the  electrostatic potential between the horizon and infinity along the symmetry axis, which is also a magnetic field line. Hence he assumed that the event horizon was an equipotential surface, citing \citet{Carter73} (later republished as \citet{Carter10}), and therefore the same potential drop existed along any magnetic field line crossing the horizon.  Based on this understanding, he envisaged an accretion by the BH of electrically charged particles which terminates when the potential drop vanishes. Based on the solution at the symmetry axis, this occurs when the electric charge of the black hole riches the critical value $q_0=2aB_0$\footnote{This value is too small to modify the BH spacetime. Hence, one may safely utilise the Kerr metric in the analysis.}. 
   
When Wald's black hole accumulates electric charge $q$, the 4-potential of its electromagnetic field becomes 
 
\beq
   U_\mu = \frac{B_0}{2}(m_\mu+2a k_\mu)  - \frac{q}{2} k_\mu \,.
\eeq
For $q=q_0$, this yields 
\beq
   U_\mu = \frac{B_0}{2}m_\mu \,,
   \label{eq:wald-q0}
\eeq
which is exactly the same as equation \eqref{eq:wald-schw} describing Wald's solution for non-rotating black hole. Based on this, \citet{KP21} concluded that the electromagnetic field is the same as in the case of a non-rotating black hole and hence the Blandford-Znajek mechanism cannot not operate.   However, the validity of this conclusion  is questionable because the spacetimes of rotating and non-rotating black holes are different, and so are the properties of their Killing vectors.

\section{Properties of the Wald solution for electrically charged black hole}
\label{sec:PWS}

\begin{figure*}
\begin{center}
\includegraphics[width=0.33\textwidth]{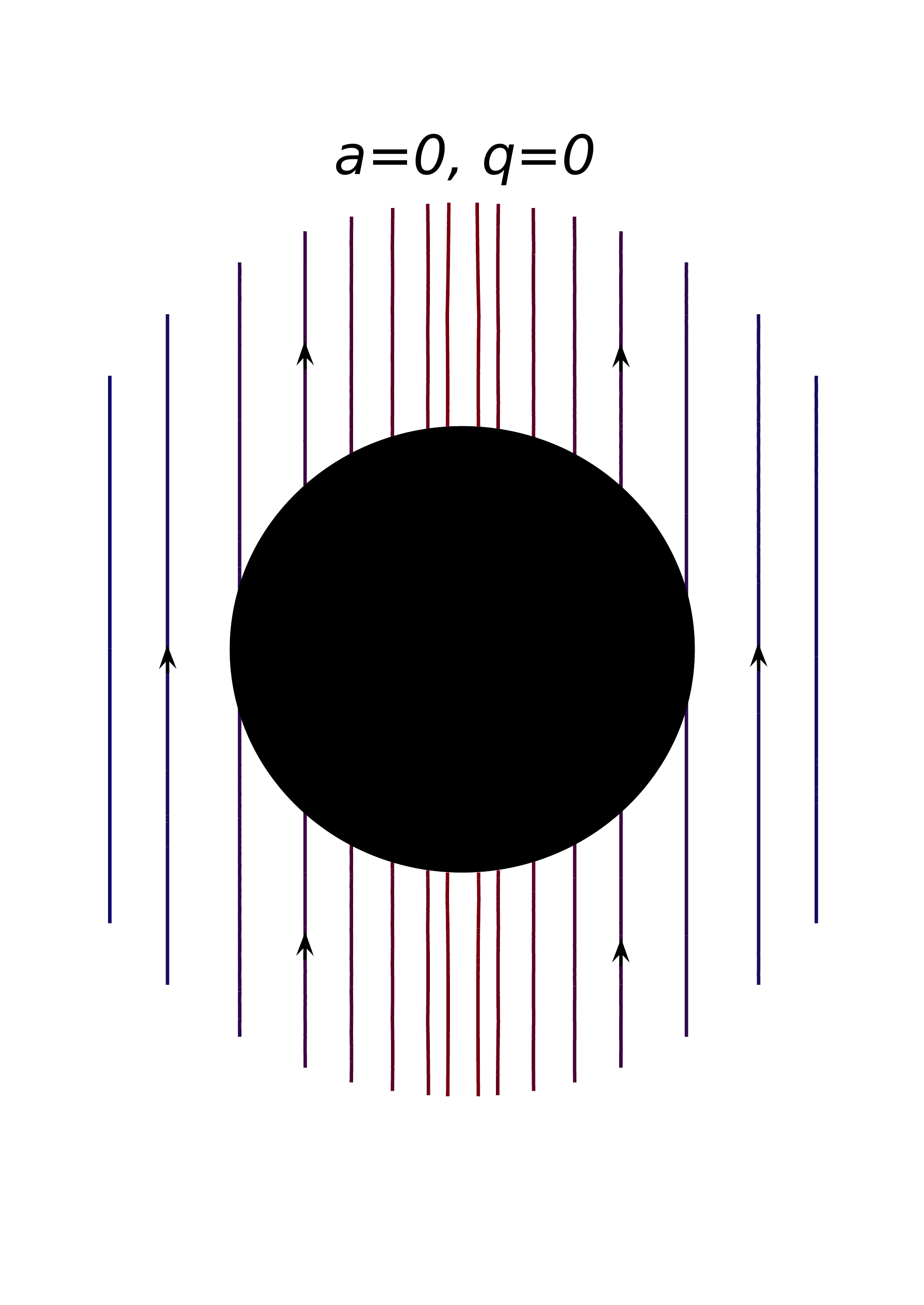}
\includegraphics[width=0.33\textwidth]{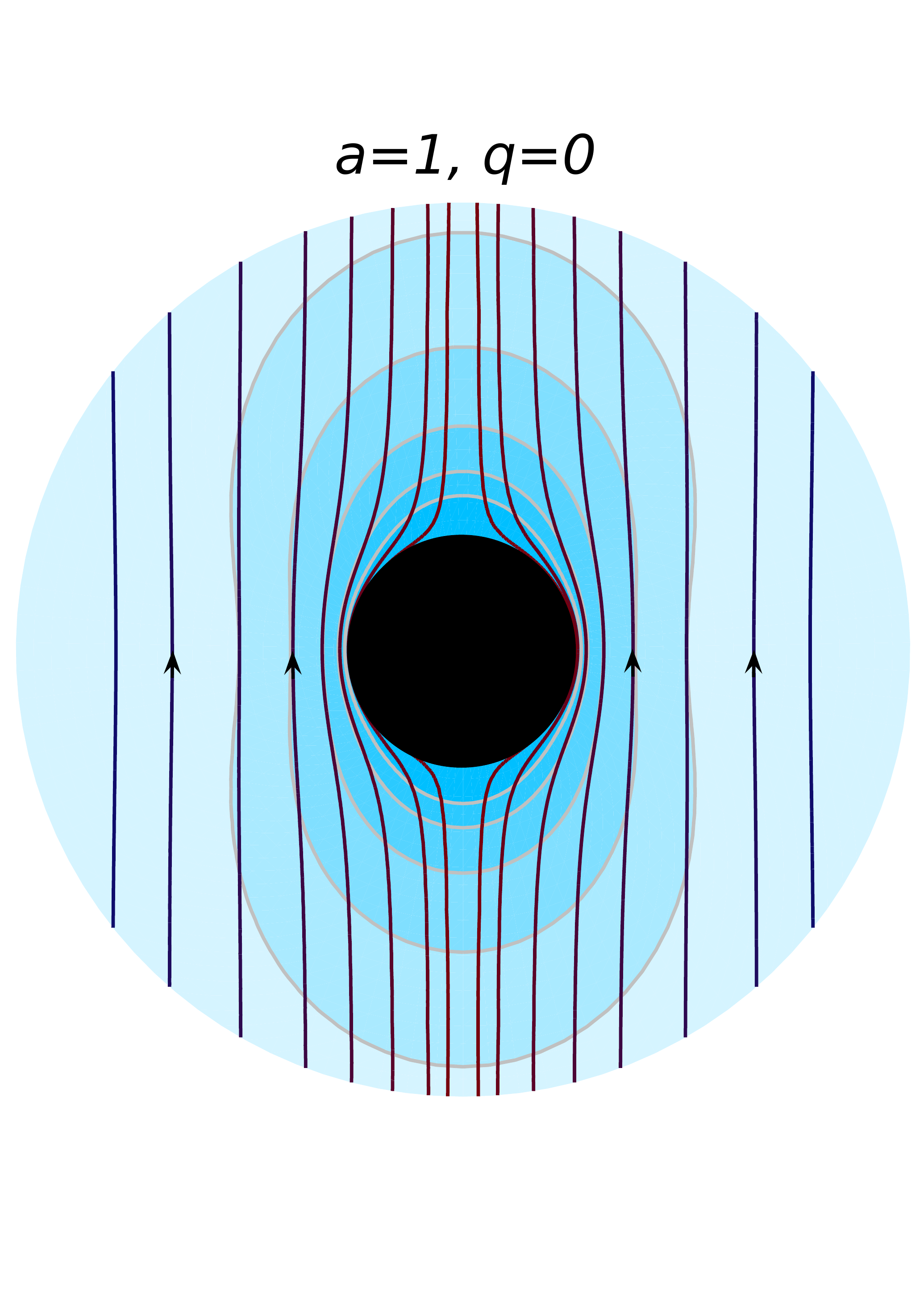}
\includegraphics[width=0.33\textwidth]{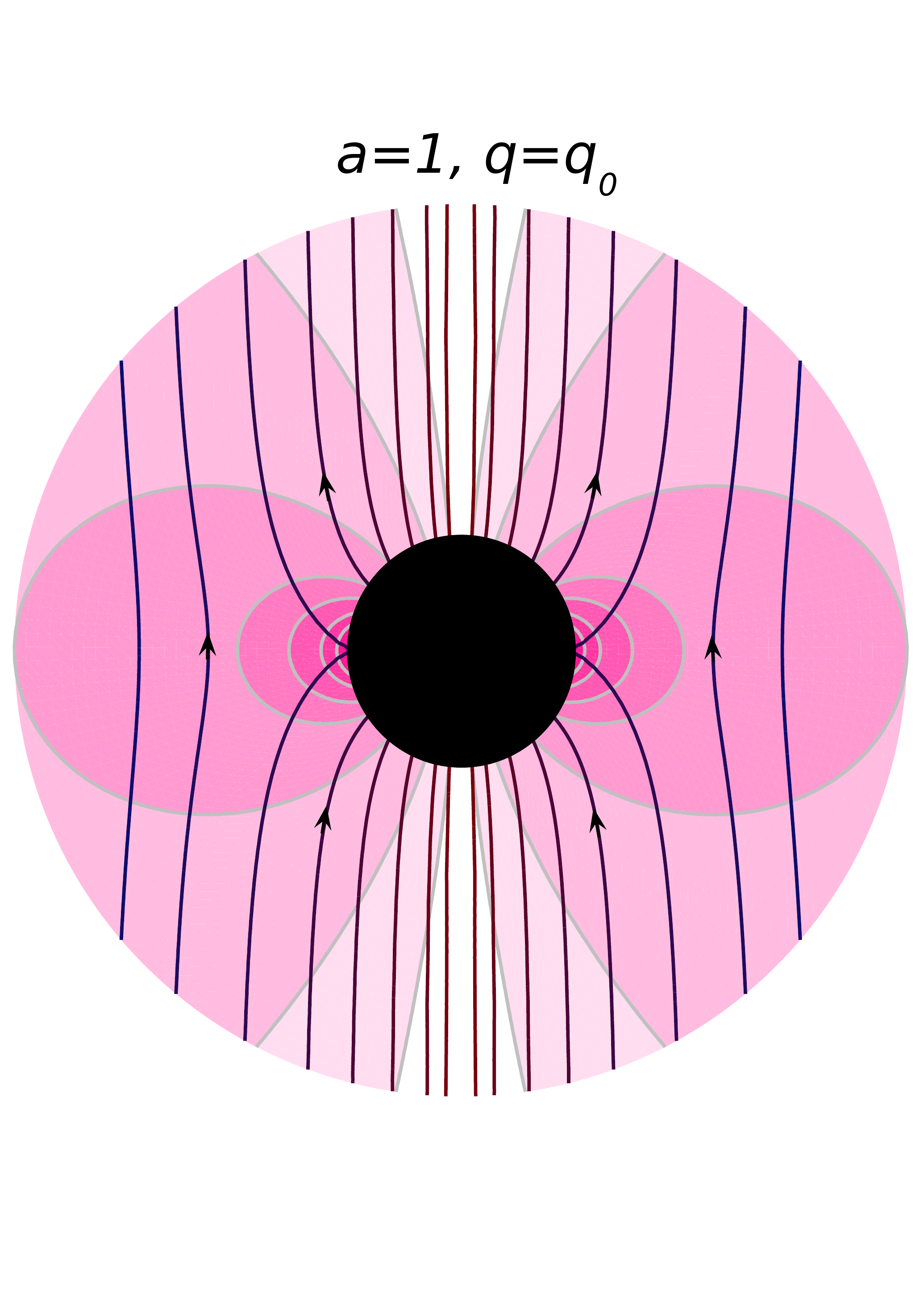}
\caption{Magnetic field lines and electrostatic potential of Wald's solutions.
{\it Left panel:}  Wald solution for $a=0$ and $q=0$. According to equation \eqref{eq:Phi_Wald},  $\Phi=0$ everywhere.    
{\it Middle panel:} Wald solution for $a=1$ and $q=0$. Contours of $(2/B_0)\Phi$ are -1.9, -1.8, -1.6, -1.3 and -1 in the order of decreasing intensity of colour/shade. 
{\it Right panel:} Wald solution for $a=1$ and $q=q_0$. Contours of $(2/B_0)\Phi$ are 0.01, 0.1, 0.5, 1.0, 1.3, 1.6, and 1.8 in the order of increasing intensity of colour/shade. 
In all the images, the magnetic field lines correspond to $(2/B_0)A_\phi=$ 0.02, 0.1,0.4,1,2,4,7, and 10. The black disk shows the event horizon. In all cases, we introduce an additive constant to our expressions for the potential so that $\Phi\to0$ as $r\to+\infty$.}
\label{fig:lines+pot}
\end{center}
\end{figure*}

The fact that in the spacetime of a rotating BH the contravariant components of the Killing vectors $\VV{m}$ and $\VV{k}$ do not depend on the spin parameter $a$ does not  mean that the same applies to their covariant components and hence to $U_\mu$. Indeed, 
\begin{align}
     m_\mu &= g_{\mu\nu}m^\nu=g_{\mu\phi} \,, \\
     k_\mu &= g_{\mu\nu}k^\nu=g_{\mu t} \,.
\end{align}
For the Boyer-Linquist coordinates $\{t,\phi,r,\theta\}$ this yields 
\begin{align}
    m_\mu &= \left(
    	-(2r/\rho^2) a\sin^2\theta\,,\,
    	\Sigma \sin^2\theta/\rho^2\,,\,
    	0\,,\,
	0
    \right) \,, \\ 
    k_\mu &=  \left(
    	-1+(2r/\rho^2) \,,\,
    	-(2r/\rho^2) a\sin^2\theta\,,\,
    	0\,,\,
	0
    \right) \,, \\ 
\end{align}
where 
\begin{align}
  \nonumber
  \rho^2 &= r^2 +a^2\cos^2\!\theta\,, \\ 
  \nonumber
   \Sigma &= (r^2+a^2)^2-a^2\Delta\sin^2\!\theta\,, \\
  \nonumber
   \Delta &= r^2+a^2-2r\,.
\end{align}
In the corresponding 3+1 splitting of $U_\mu$, the electrostatic potential $\Phi=-U_t$ and the magnetic vector potential $A_i=U_i$, with $i=\phi,r,\theta$. Hence  

\beq
      \Phi = \frac{B_0}{2} \left( \frac{2r}{\rho^2} a\sin^2\theta  +2a\left( 1-\frac{2r}{\rho^2} \right) \right) \,,
      \label{eq:Phi_Wald}
\eeq

\beq
      A_\phi = \frac{B_0}{2} \left(\frac{\Sigma}{\rho^2} \sin^2\theta +2a\left( -\frac{2r}{\rho^2} a\sin^2\theta\right) \right) \,.
       \label{eq:A_Wald}
\eeq
for the Wald solution with $q=0$, and 

\beq
      \Phi = \frac{B_0}{2} \frac{2r}{\rho^2} a\sin^2\theta 
      \label{eq:Phi_q0}
\eeq

\beq
      A_\phi = \frac{B_0}{2}  \frac{\Sigma}{\rho^2} \sin^2\theta  
      \label{eq:A_q0}
\eeq
for the solution with $q=q_0$. The component $A_\phi$ is particularly relevant for our analysis because it is invariant along the magnetic field lines of axisymmetric solutions. Using these equations, we can explore the differences in the properties of the Wald solutions for charged and uncharged black holes.

\begin{figure*}
\begin{center}
\includegraphics[width=0.4\textwidth]{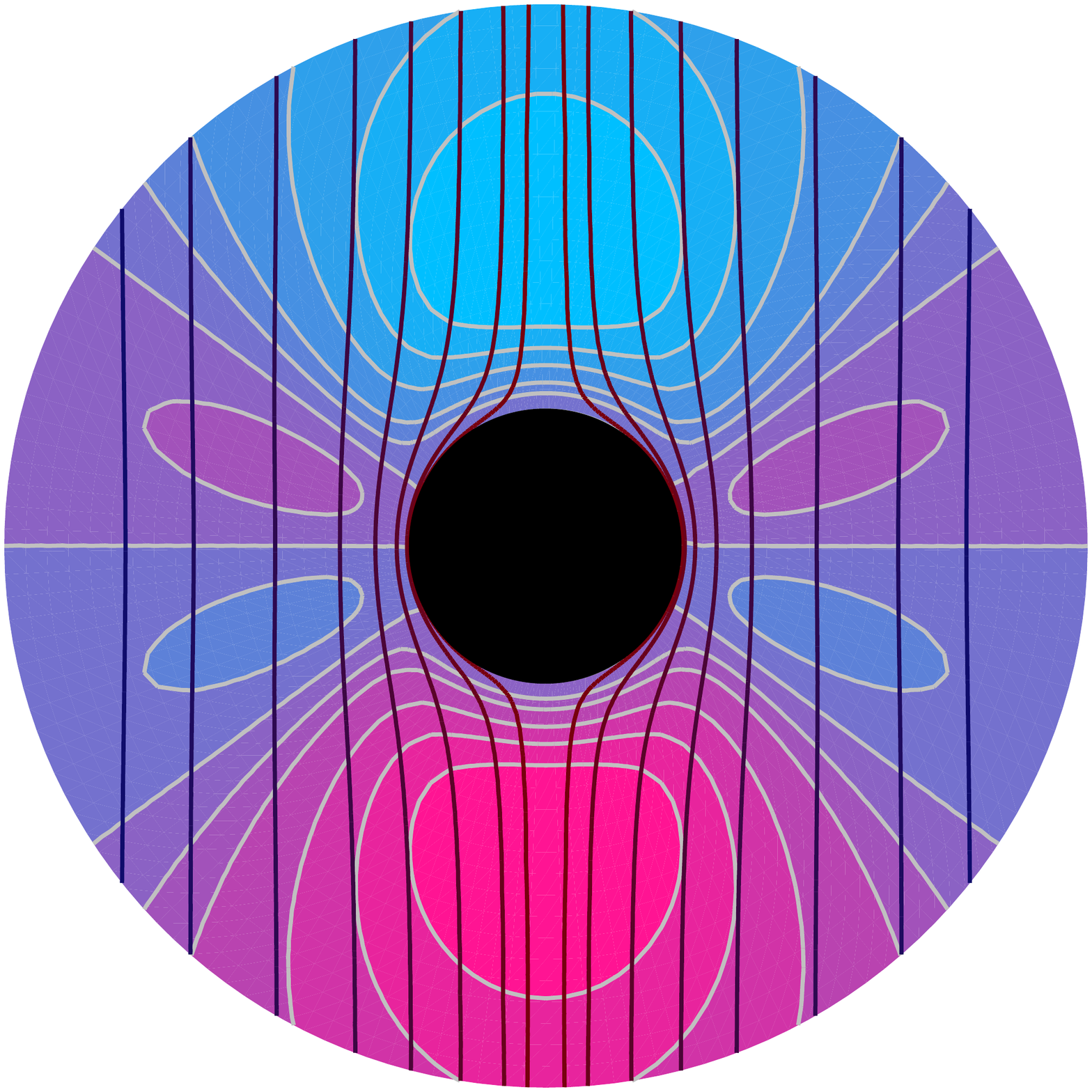}
\includegraphics[width=0.4\textwidth]{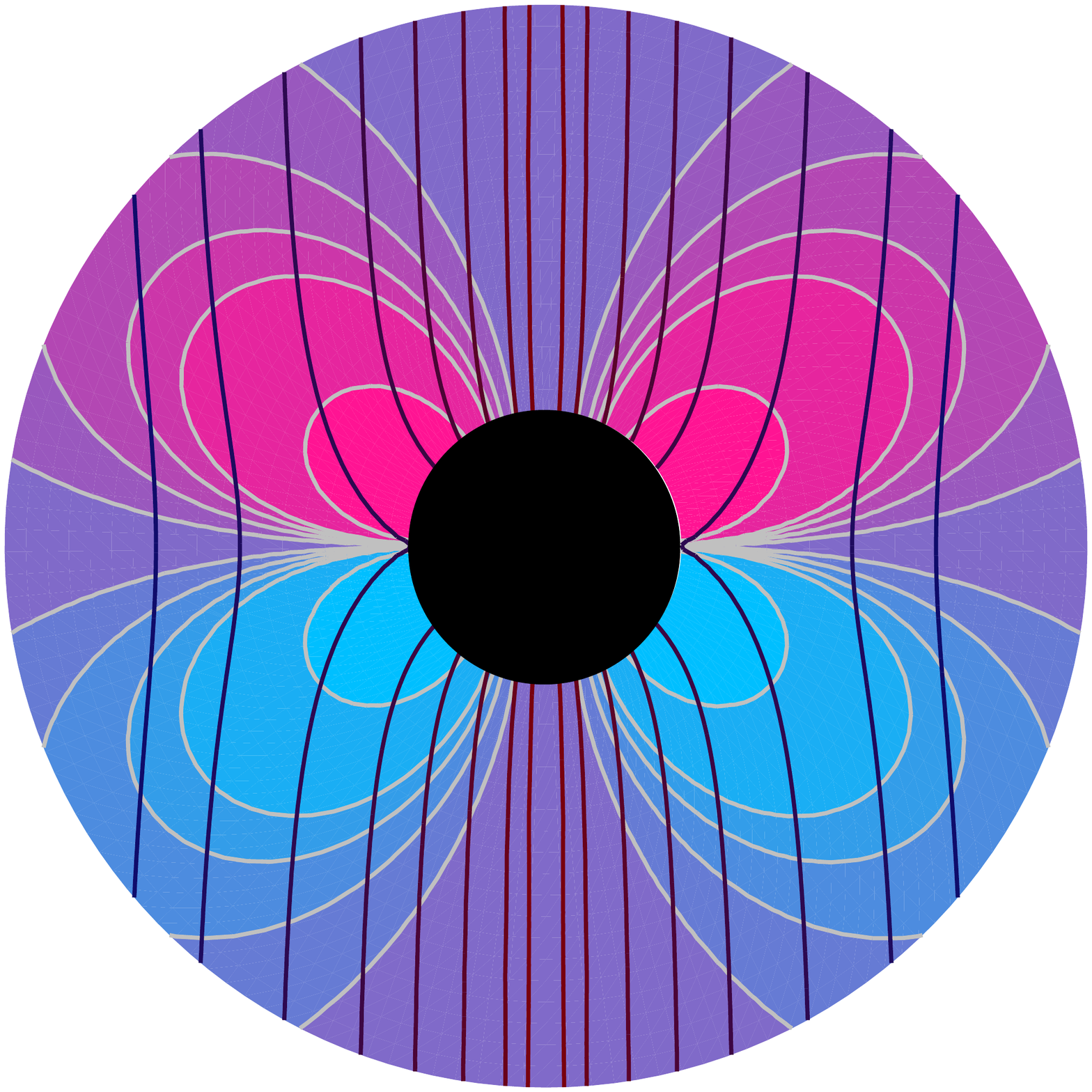}
\caption{Magnetic field lines and the invariant $\Fs_{\mu\nu}F^{\mu\nu}$ of the electromagnetic field for a rotating BH. 
{\it Left panel:} BH with $a=1$ and $q=0$. Contours of $(4/B_0^2)\Fs_{\mu\nu}F^{\mu\nu}$ are -4,-3,-2,-1,0,1,2,3,4 in the order from blue to pink colour. 
{\it Right panel:} BH with $a=1$ and $q=q_0$. Contours of $(4/B_0^2)\Fs_{\mu\nu}F^{\mu\nu}$ are -10,-4,-3,-2,-1,1,2,3,4, and 10 in the order from blue to pink colour. In both the images, the magnetic field lines correspond to $(2/B_0)A_\phi=$ 0.02, 0.1,0.4,1,2,4,7, and 10. The black disk shows the event horizon.}
\label{fig:BE+lines}
\end{center}
\end{figure*}

\begin{figure*}
\begin{center}
\includegraphics[width=0.4\textwidth]{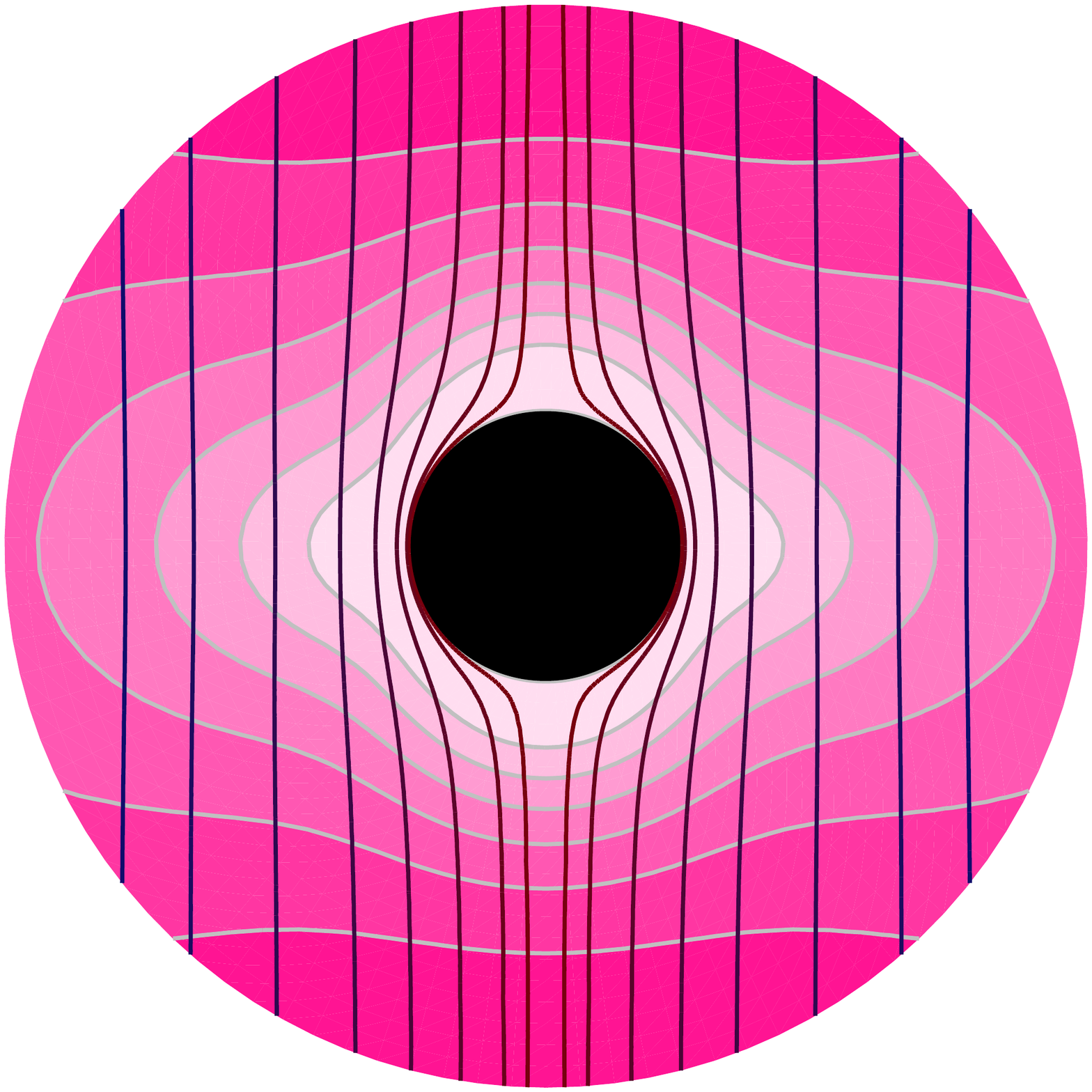}
\includegraphics[width=0.4\textwidth]{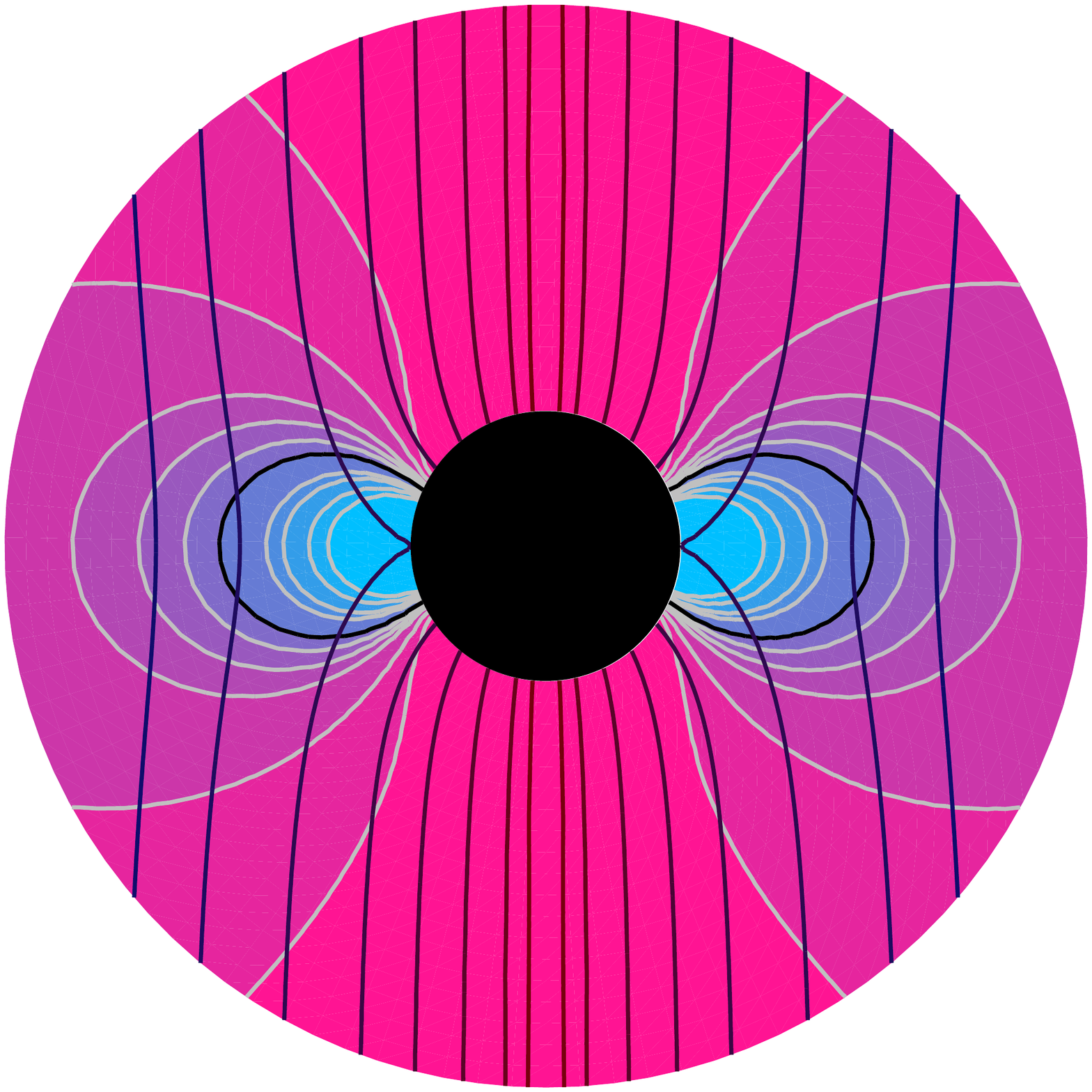}
\caption{Magnetic field lines and the invariant $F_{\mu\nu}F^{\mu\nu}$ of the electromagnetic field for a rotating BH. 
{\it Left panel:} BH with $a=1$ and $q=0$. Contours of $(4/B_0^2)F_{\mu\nu}F^{\mu\nu}$  are 0,1,2,3,4,5, and 6 in the order of increasing intensity of colour/shade. 
{\it Right panel:} BH with $a=1$ and $q_0$. Contours of $(4/B_0^2)F_{\mu\nu}F^{\mu\nu}$ 
are -7,-5,-3,-2,0,1,2,3,5, and 7 in the order from blue to pink colour.  The black contour corresponds to $F_{\mu\nu}F^{\mu\nu}=0$.
In both the images, the magnetic field lines correspond to $(2/B_0)A_\phi=$ 0.02, 0.1,0.4,1,2,4,7, and 10. The black disk shows the event horizon.}
\label{fig:B2-E2+lines}
\end{center}
\end{figure*}

Figure \ref{fig:lines+pot} shows the magnetic field lines and $\Phi$  for 1) $a=0$, $q=0$, 2)  $a=1$, $q=0$, and 3) $a=1$, $q=q_0$.  ( The case of maximal rotation was chosen simply to increase the contrast between the solutions. Qualitatively, the solutions for $0<a<1$ are the same.)  One can see that, in contrast to the Wald solution for a non-rotating BH, the solution for a rotating BH with $q=q_0$ does exhibit a drop of the electrostatic potential along all the magnetic field lines but the one coinciding with the symmetry axis.   

The relativistic Lagrangian for a particle of rest mass $m_p$, electric charge $q_p$, and 4-velocity $u^\nu$ is
\beq
  L =\frac{1}{2}m_p g_{\nu\mu}u^\nu u^\mu +e_p U_\nu u^\nu \,.
\eeq
For a time-independent space-time and electromagnetic field, the Euler-Lagrange equation with this Lagrangian yield the integral of motion $e_p^\infty = -m_p u_t +q_p\Phi$, the total redshifted energy of the particle. Hence, the variation of $\Phi$ along the magnetic field lines in the solution with $a=1$ and $q=q_0$ implies that charges particles can be accelerated by its electric field\footnote{\citet{Wald74} used exactly the same analysis, but for some reason applied it only to the symmetry axis.}.

Alternatively, the same conclusion can be made by inspecting the 4-scalars $\Fs_{\mu\nu}F^{\mu\nu}$ and  $F_{\mu\nu}F^{\mu\nu}$, where $F_{\mu,\nu} = U_{\nu,\mu} - U_{\mu,\nu}$ is the Maxwell tensor of the electromagnetic field and $\Fs_{\mu\nu}=e_{\mu\nu\alpha\beta}F^{\alpha\beta}$ is its Hodge dual tensor. In any local inertial frame,  $$
\Fs_{\mu\nu}F^{\mu\nu}=4\spr{\cE}{\cB} \,,
$$ 
$$
F_{\mu\nu}F^{\mu\nu}=2(\cB^2-\cE^2)\,,
$$ 
where $\cE$ and $\cB$ are the electric and magnetic fields as measured in this frame, respectively. Hence, $\Fs_{\mu\nu}F^{\mu\nu}\not=0$ implies acceleration of electrically-charged particles along the magnetic field lines. If $\Fs_{\mu\nu}F^{\mu\nu}=0$ but $F_{\mu\nu} F^{\mu\nu}<0$ the particles will be accelerated perpendicular to the magnetic field lines. 

Figure \ref{fig:BE+lines} shows the distributions of  $\Fs_{\mu\nu}F^{\mu\nu}$ for BHs with $q=0$ and $q=q_0$ ($a=1$ in both the cases). For $q=0$, the value of $\spr{\cE}{\cB}$ vanishes only in the equatorial plane and at $\theta\approx 90^\circ\pm29^\circ$ (This agrees with the analysis by \citet{King75}). Its magnitude has broad tall peaks in the polar regions and less pronounced peaks at about $15^\circ$ from the equator. In contrast,  for $q=q_0$, the value of $\spr{\cE}{\cB}$ vanishes at the equatorial plane and the symmetry axis,  and peaks at mid-latitudes (see also \citet{Levin18}).  

Figure \ref{fig:B2-E2+lines} shows the corresponding distributions of  $F_{\mu\nu}F^{\mu\nu}$.  For $q=0$, $\cB^2>\cE^2$ everywhere outside of the event horizon. In contrast, for $q=q_0$, $\cB^2<\cE^2$ inside the toroidal region symmetric about the equatorial plane. In figure \ref{fig:B2-E2+lines}, the outer boundary of this region is indicated by the black contour of the distribution of $F_{\mu\nu}F^{\mu\nu}$. Thus, in the charged case, charged particles can be accelerated across the magnetic field lines in an  equatorial region close to the event horizon (This region extends slightly beyond the BH ergosphere.)        

These results demonstrate convincingly that charging of BHs up to the ``critical'' electric charge $q_0=2aB_0$ does not remove the possibility of particle acceleration, copious pair creation and vacuum breakdown, and hence the possibility of operation for the  Blandford-Znajek mechanism.   

\begin{figure}
\begin{center}
\includegraphics[width=\columnwidth]{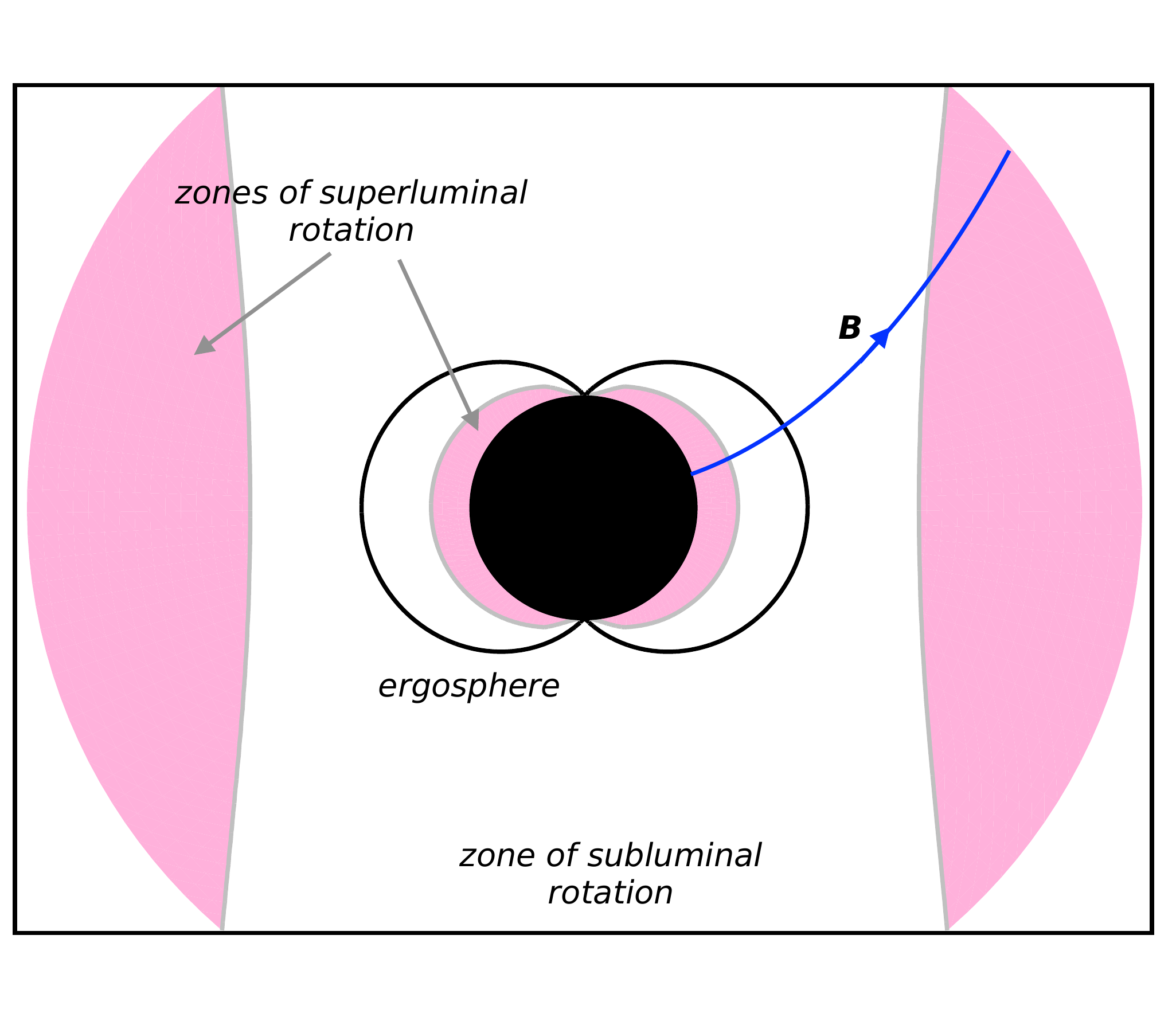}
\caption{Superluminal and subluminal zones for rotation with $\Omega=0.5 \Omega_h$, where $\Omega_h=a/(r_+^2+a^2)$ is the angular velocity of the black hole with $a=1$.  }
\label{fig:lsurf}
\end{center}
\end{figure}

\section{The general case of a stationary axisymmetric magnetosphere}
\label{sec:GC}

Since the Wald solution is highly specific, it is not clear how general the conclusions drawn from its analysis are. The BH electric charge may differ from $q_0$, the electric charge may be accumulated in clouds around the BH, and the magnetic field may have a very different configuration. This issue has been addressed in \citet{ssk-ebh}, where it is shown that the Wald solution is just an example of a general rule.  Here, we only outline the arguments  and refer interested readers to the original.      

Let us allow not only the black hole to be electrically charged but also allow electric charge to be distributed around the black hole.  This spatially distributed charge may be associated with an azimuthal electric current, but the poloidal current is not allowed because this is a key property of "live" magnetospheres, where the rotational energy of black hole is extracted electromagnetically.

In a "dead" magnetosphere, the electric field is completely screened in the sense that 
\begin{equation}
  \Fs_{\mu\nu}F^{\mu\nu} = 0,     
\label{con1}
\end{equation}
and
\begin{equation}
  F_{\mu\nu}F^{\mu\nu} > 0.     
\label{con2}
\end{equation}
In order to show that this is impossible under the described conditions, let us assume that the first of these conditions is satisfied everywhere and hence show that this leads to breakdown of the second condition near the BH.

The condition \eqref{con1} means that the parallel component of the electric field is screened, and for an axisymmetric configuration this implies that  
\begin{equation}
   \bE =-\Omega (\vpr{m}{B})\,,
\end{equation}
where $E_i=F_{it}$, $B^i=(1/2)e^{ijk}F_{jk}$ and $\vv{m}$ is the spatial component of the Killing vector $\VV{m}=(0,\vv{m})$ (
The same result holds when the magnetic field is frozen in plasma rotating with the angular velocity $\Omega$.).  In a stationary case, $\Omega$ is constant along the magnetic field lines and for this reason can be interpreted as the angular velocity of these lines. For such an electric field, 
\begin{equation}
  F_{\mu\nu}F^{\mu\nu} = - \frac{2}{\alpha^2} B^2 f(\Omega,r,\theta),     
\label{o4}
\end{equation}
where
\begin{equation}
 f(\Omega,r,\theta)=g_{\phi\phi}\Omega^2+2g_{t\phi}\Omega+g_{tt}   
\label{o5}
\end{equation}
is called the light surface function.  For $f<0$, a point rotating with the angular velocity $\Omega$ has a time-like worldline (subluminal rotation), and for $f>0$ its world line is space-like (superluminal rotation). Equation \eqref{o4} shows that for a point of the magnetic field line the subluminal rotation  implies $F_{\mu\nu}F^{\mu\nu}>0$ ($\cE<\cB$), and the superluminal rotation implies $F_{\mu\nu}F^{\mu\nu}<0$ ($\cE>\cB$). 

For any $0<\Omega<\Omega_h$, where $\Omega_h=a/(r_+^2+a^2)$ is the angular velocity of the BH, the space around BH includes an inner region of superluminal rotation, an outer region of superluminal rotation, and a region of subluminal rotation in between them \citep{ssk-ebh}. The surfaces separating these regions are called light surfaces. For $r\gg r_+$, the outer light surface has the shape of a cylinder with the radius $\varpi=1/\Omega$. The inner light surface is always outside of the event horizon, touching it at $\theta=0,\pi$, and approaches the ergosphere as $\Omega\to 0$ (see figure \ref{fig:lsurf}).  For $\Omega>\Omega_h$ the inner light surface disappears but the outer light surface crosses the event horizon. Thus, there always exists a region near the event horizon where condition \eqref{con2} is not satisfied. 

It turns out that both the screening conditions can be satisfied everywhere if one allows poloidal electric current through the magnetosphere \citep{ssk-ebh}. However, this implies live magnetosphere with operating BZ mechanism.   In order to sustain such currents charged particles have to be constantly created in-situ, for example in the potential gaps of the magnetosphere \citep[e.g][]{Beskin92,Crinquand20}.

\begin{figure}
\begin{center}
\includegraphics[width=0.8\columnwidth]{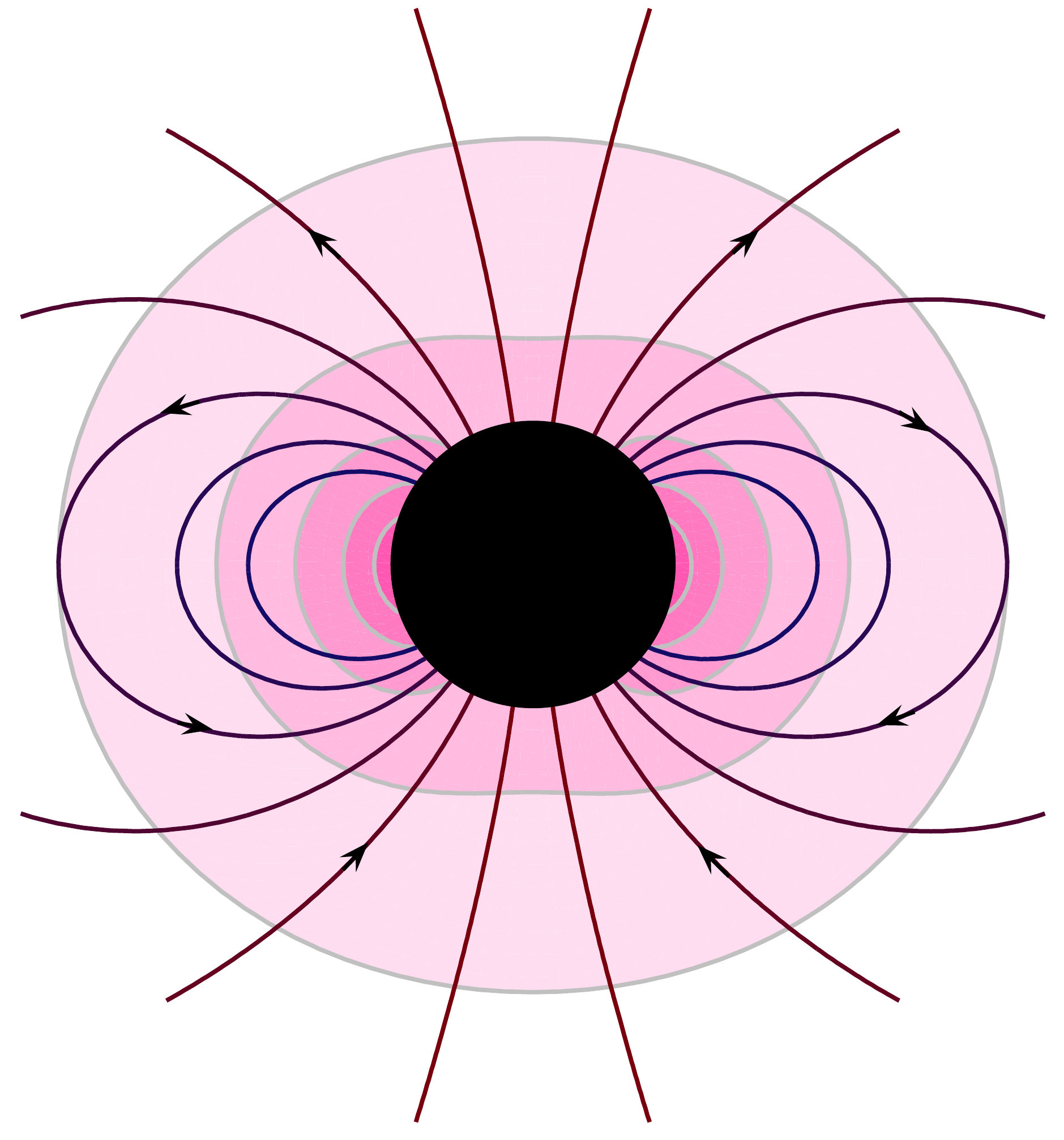}
\caption{Magnetic field lines and electrostatic potential of Wald's solution
for an electrically-charged BH with $a=1$ and $B_0=0$. Contours of $(2/q)\,\Phi$ are 0.6, 0.9, 1.2, 1.5 and 1.8 in the order of decreasing intensity of colour/shade. The magnetic field lines correspond to $(2/q)A_\phi=$ 0.02, 0.2,0.4,0.6,0.8, and 1.0}
\label{fig:isolated}
\end{center}
\end{figure}

\begin{figure*}
\begin{center}
\includegraphics[width=0.4\textwidth]{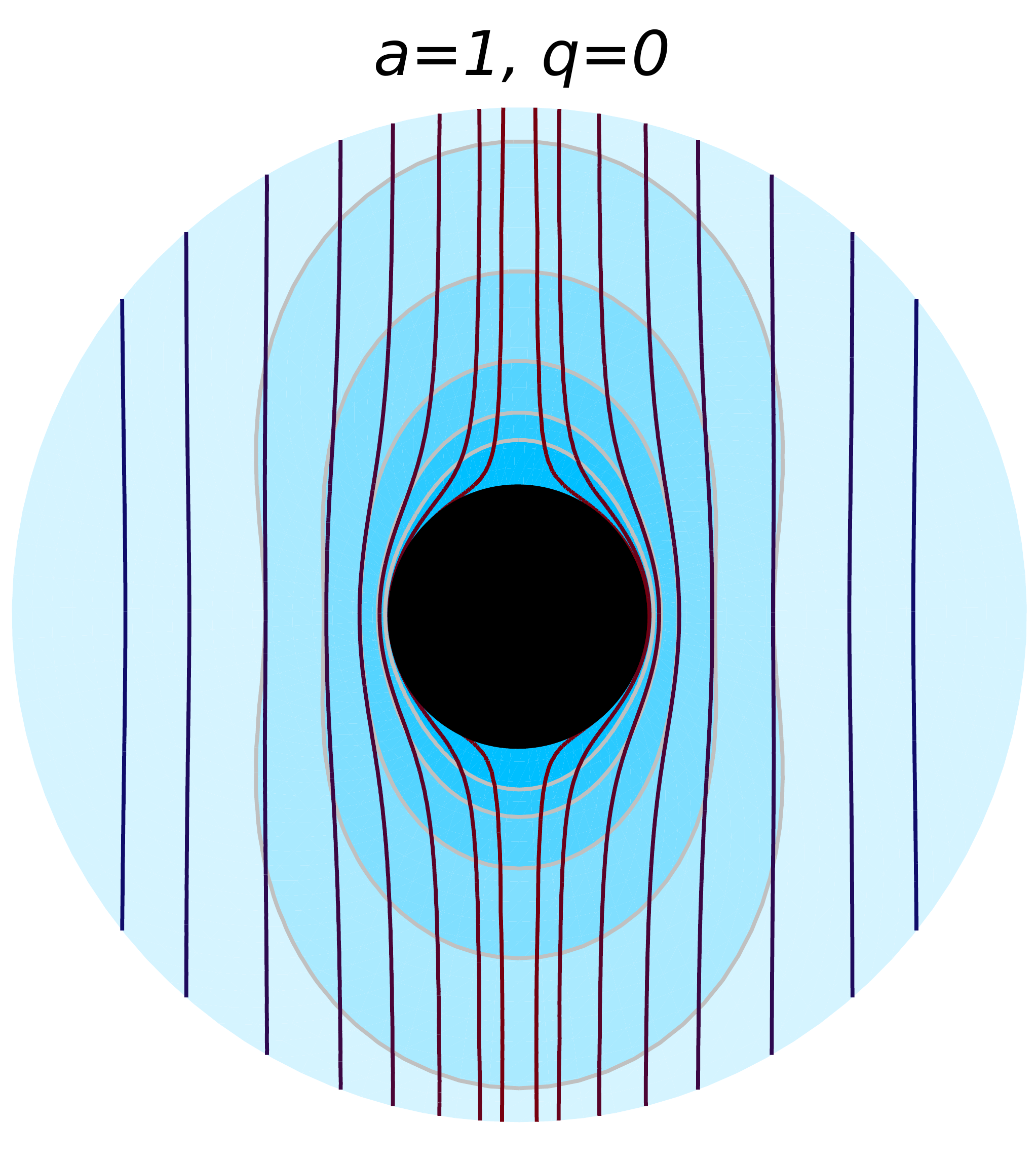}
\includegraphics[width=0.4\textwidth]{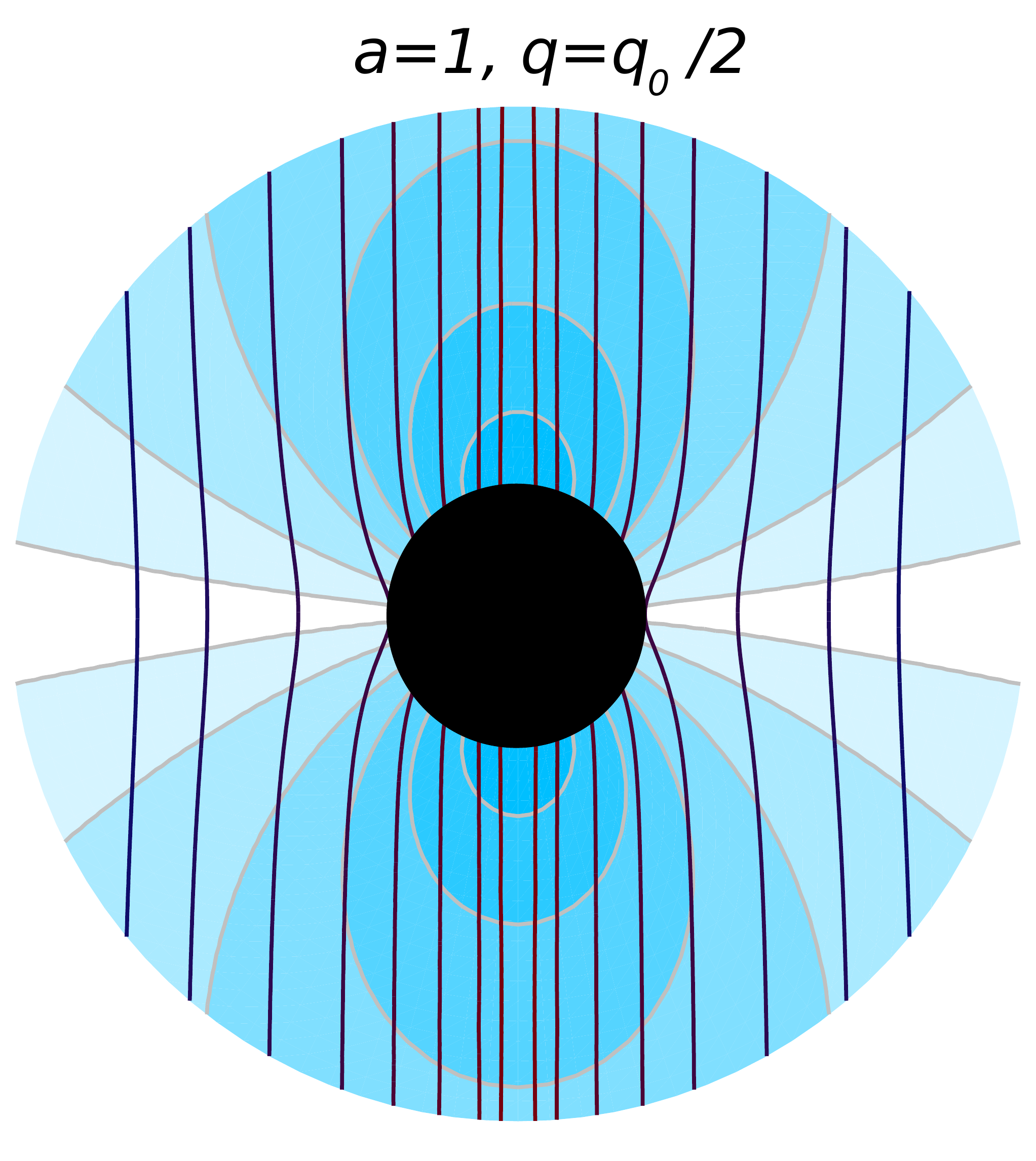}
\includegraphics[width=0.4\textwidth]{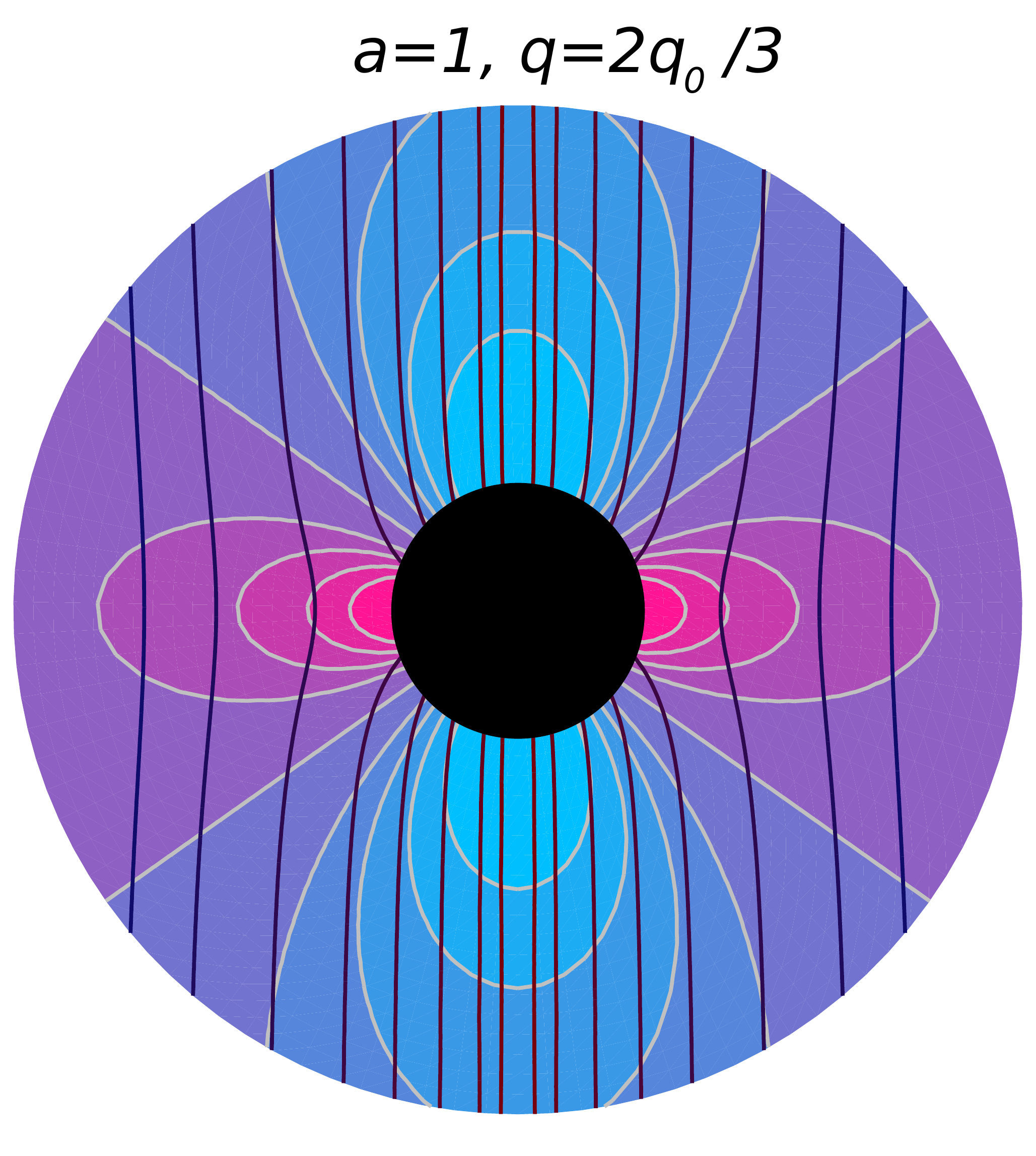}
\includegraphics[width=0.4\textwidth]{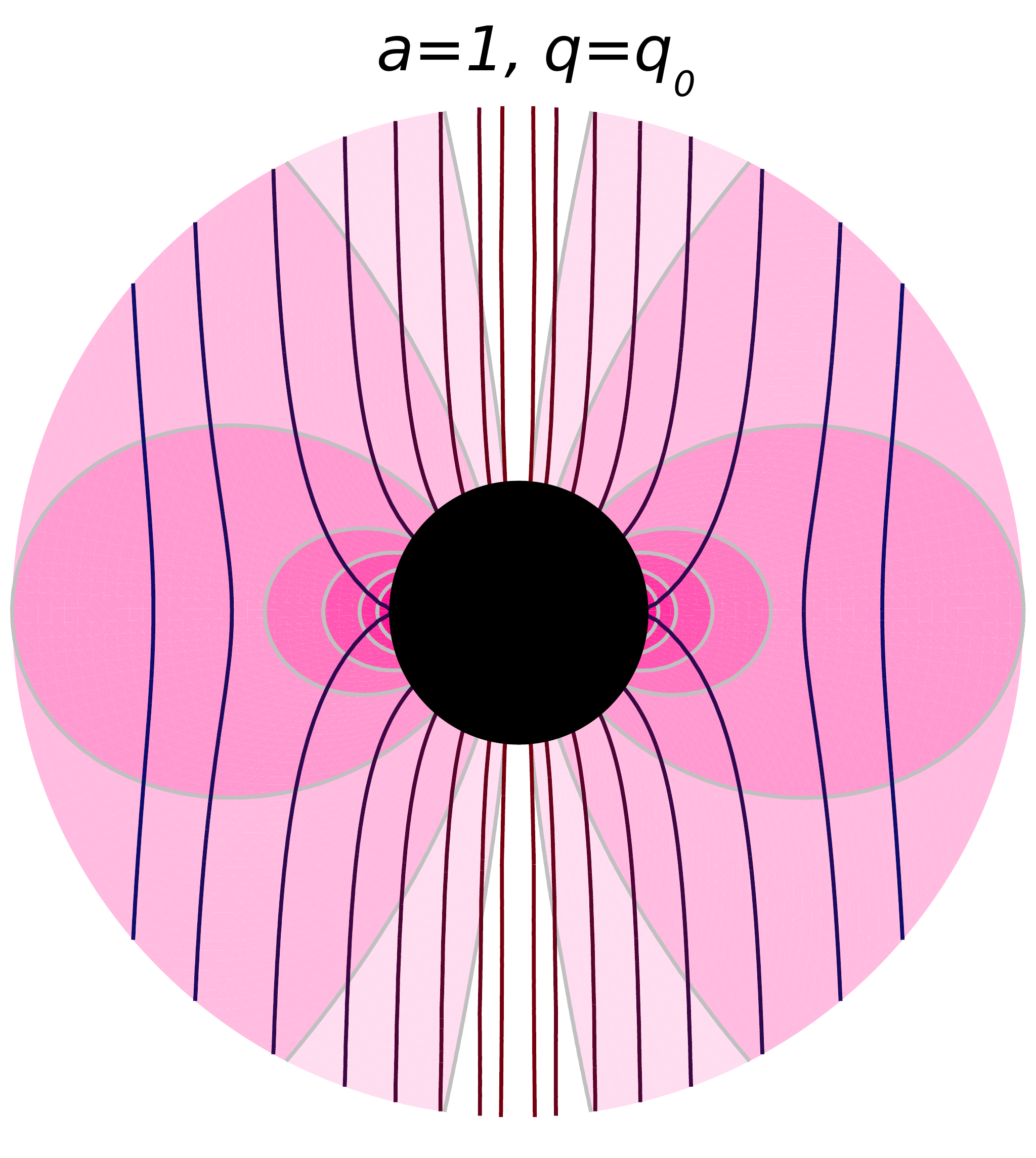}
\caption{Dependence of the magnetic field lines and electrostatic potential on the BH electric charge.
{\it Top-left panel:}  Wald's solution for $a=1$ and $q=0$. Contours of $(2/B_0)\,\Phi$ are -1.9, -1.8, -1.6, -1.3 and -1 in the order of decreasing intensity of colour/shade. 
{\it Top-right panel:}  Wald's solution for $a=1$ and $q=0$. Contours of $(2/B_0)\,\Phi$ are -0.9, -0.7, -0.5, -0.3, -0.1 and -0.01 in the order of decreasing intensity of colour/shade. 
{\it Bottom-left panel:} Wald's solution for $a=1$ and $q=q_0$. Contours of $(2/B_0)\,\Phi$ are -0.5,-0.4,-0.3,-0.2,0,0.2,0.3,0.4,0.5 in the order of increasing intensity of colour/shade. 
{\it Bottom-right panel:} Wald's solution for $a=1$ and $q=q_0$. Contours of $(2/B_0)\,\Phi$ are 0.01, 0.1, 0.5, 1.0, 1.3, 1.6, and 1.8 in the order of increasing intensity of colour/shade. 
In all the images, the magnetic field lines correspond to $(2/B_0)\,A_\phi=$ 0.02, 0.1,0.4,1,2,4,7, and 10. The black disk shows the event horizon. In all cases, we introduce an additive constant to our expressions for the potential so that $\Phi\to0$ as $r\to+\infty$.}
\label{fig:var-q}
\end{center}
\end{figure*}

\section{Discussion}
 \label{sec:Disc}

Our study does not answer the key question raised in \citet{Wald74} of whether rotating BH can naturally acquire electric charge via interacting with the surrounding magnetised plasma.  However, in combination with other studies, it suggests
that this may be the case. 

In particular, figure \ref{fig:lines+pot} shows a rather interesting difference in the magnetic field distribution in the solutions for $q=0$ and $q=q_0$. In the solution for $q=0$, the magnetic field lines are expelled from the BH horizon (This immediately follows from equation \eqref{eq:A_Wald} which yields $A_\phi(r_+)=0$ for $a=1$.)  This is the well-known Meissner effect of BH electrodynamics \citep{King75}\footnote{On a separate note, it is worth pointing out that the fact that the magnetic field is completely expelled from the event horizon of the maximal-rotating  BH ($a=1$) shows that it is incorrect to interpret the generated electric field as the result of the electric charge separation on the stretched horizon of the membrane paradigm. Instead, the electric field is gravitationally induced and should be attributed to the extreme differential rotation of the space around the BH, manifesting itself in the dragging of local inertial frames.}.

In contrast, for $q=q_0$, the magnetic field lines are pulled onto the horizon. A similar effect was observed in the numerical force-free solutions to the Wald problem \citep{ssk-meissner} and in the recent PIC simulations \citep{Parfrey19}.  The reason for this is easy to understand if one considers the structure of the electromagnetic field of the BH with $B_0=0$ and $q\not=0$. According to \citet{Wald74}, its potential is simply 
$$
  U_\mu= -\frac{q}{2} k_\mu \,.
$$ 
Figure \ref{fig:isolated} shows the corresponding field of the electrostatic potential and the magnetic field lines for the case with the spin parameter $a=1$. One can see that the magnetic field is dipolar and all its magnetic field lines penetrate the event horizon. When this solution is added to the one for the uncharged BH, some magnetic field lines of the resultant total magnetic field also become connected to the event horizon.  

\citet{ssk-meissner} attributed the effect of pulling the magnetic field lines back onto the BH event horizon in their force-free simulations  to the pinching of the axial poloidal electric current (see also \citet{ssk-ebh,NC14,PY16,Pan18}), but our results show that a finite BH charge may well be another reason. The latest PIC simulations for the case with an external magnetic field of paraboloidal geometry show that BHs do indeed acquire electric charge of the order of $aB_0$, where $B_0$ is the magnetic field strength in the vicinity of the BH (Benjamin Crinquand, private communication). It would be interesting to rerun or (reanalyse) the force-free and PIC simulations for the Wald problem and to measure the accumulated charge. Hence one can compare the magnetic configuration observed in the simulations to that of the vacuum solution with the same BH charge and determine the importance of the BH charge.    

With regard to computer simulations, \citet{KP21} criticised their setup for not being careful with the inner boundary conditions.  In particular, they claimed that the removal of PIC particles after they have crossed the event horizon \citep{Parfrey19,Crinquand20} could prevent BH from acquisition of electric charge. However, this argument ignores the basic causal structure of the BH spacetime.   Once the inner boundary is placed inside the event horizon its conditions cannot influence the solution in the exterior to the horizon. The same applies to the removal of PIC particles. This may upset the interior solution, but the imprint left by the removed particle in the exterior solution remains unaffected\footnote{The possibility of placing the inner boundary inside the event horizon is the key benefit of using the Kerr-Schild coordinate system in computer simulations.}.  A similar argument can be made with regard to force-free and RMHD simulations. There is no reason to expect the total electric current through the event horizon to be exactly zero, especially in the initial phase where the solution may be far from a steady-state. During this phase, the BH may accumulate an electric charge which will be detectable via the structure of the steady-state solution, or quasi-steady-state solution, exterior to the event horizon. For example, \citet{Levin18} have shown that in the split-monopole force-free solution by \citet{BZ-77} the BH hole has a non-vanishing electric charge.

Since Wald's conclusion that the event horizon of a BH with the electric charge  $q_0=2aB_0$ is equipotential is incorrect, one would not expect the BH  charge to saturate exactly at $q_0$.  Indeed, the comparison of the top-left and bottom-right panels of figure \ref{fig:var-q} shows that for $q=q_0$ the polarity of the electric field is opposite to that for $q=0$, and hence the BH attracts only charges of the opposite sign. Hence, if the equilibrium is reached, the corresponding BH charge must be strictly below the value of $q_0$. Moreover, in this equilibrium state the BH must attract positive charges along some magnetic field lines and negative charges along the others, with the total electric current through the event horizon vanishing. 

The possibility of such state is demonstrated by the solutions corresponding to $q=q_0/2$ and $q=2/3q_0$ (see the top-right and bottom-left panels of figure \ref{fig:var-q} ). Whereas in the case with $q=0$ the electric field of BH pulls in positive charges and repels negative charges along all magnetic field lines penetrating the event horizon, the case with $q=q_0/2$ is more complicated. Sufficiently high above the event horizon, everything looks the same as in the case with $q=0$. However near the event horizon, the situation is reversed for the magnetic field lines entering the horizon near its equator. In this region,       
the BH electric field now repels positive charges and attracts negative charges. The bifurcation occurs at the point where the magnetic field line is tangent to the contour of the electrostatic potential. Around the locus of such points, one may expect to find a ``cloud'' of positive charges.  The solution for $q=2/3q_0$ is qualitatively the same, but its event horizon is characterised by a smaller polar region attracting positive charges and a wider equatorial band attracting negative charges. Along the magnetic field lines passing by the event horizon, negative charges are pulled towards the equatorial plane, where a disk of negative charges may develop. 

The electromagnetic field due to the charge clouds could result in significant deviation from the Wald solution. Moreover, this is only one of the factors that may influence the value of the equilibrium charge $q_{eq}$ of astrophysical BHs. Another factor is the way charged particles are introduced into the magnetosphere. For example, if the particles are introduced only near the symmetry axis, one expects $q_{eq}\approx q_0$. A much lower value is expected if they are introduced mainly at the equatorial plane.   For an accreting BH, the external magnetic field may also be quite different from the uniform field assumed by Wald, even close to the BH (for example, see figure 3 in \citet{ssk-meissner}).  Thus, contrary to the optimistic conclusion by \citet{Wald74}, the actual value of the electric charge accumulated by astrophysical BHs is difficult to predict theoretically. Hopefully, future computer simulations will clarify this issue.

\section{Conclusion}
 \label{sec:Con}

Whatever is the value of the electric charge accumulated by a rotating BH, it cannot not invalidate the BZ-mechanism.  It may change some details of the magnetospheric dynamics but it cannot negate the gravitationally induced electric field and hence cannot result in a dead magnetosphere. Our results agree with the Wald's conclusion that, BHs may acquire electric charge when placed into magnetic field of external origin. However, the value  $q_0=2aB_0$ given by Wald is only an upper limit, and the actual value of the BH charge cannot be given without specifying the way charged particles are introduced into the BH magnetosphere and accounting for their dynamics.   For more realistic astrophysical settings, deviation of the external magnetic field from the uniform configuration of the Wald problem can also be important.  The BH electric charge, and the corresponding dipolar magnetic field, may still be strong enough to alter the properties of the BH magnetosphere. In particular, it may cancel the so-call Meissner effect of vacuum solutions. This could be the main reason for the cancelling observed in the previous  force-free and RMHD simulations.

\section*{Acknowledgments}
All non-trivial calculations of this study were carried out with the software package {\it Maple} (Maple is a trademark of Waterloo Maple Inc.).   I would like to thank  Charles Gammie, Benjamin Crinquand, Andrew King, and Jim Pringle for helpful discussions and useful comments.



\bibliographystyle{mnras}
\bibliography{../BibFiles/bholes,../BibFiles/jets,../BibFiles/komissarov}

\begin{thebibliography}{16}
\expandafter\ifx\csname natexlab\endcsname\relax\def\natexlab#1{#1}\fi

\bibitem[{Beskin} et~al.(1992){Beskin}, {Istomin} \& {Parev}]{Beskin92}
{Beskin} V.~S., {Istomin} Y.~N., {Parev} V.~I., 1992, \sovast, 36, 642

\bibitem[{Blandford} \& {Znajek}(1977)]{BZ-77}
{Blandford} R.~D., {Znajek} R.~L., 1977, \mnras, 179, 433

\bibitem[{Carter}(1973)]{Carter73}
{Carter} B., 1973, in { Black Holes (Les Astres Occlus)\/}, edited by
  C.~{DeWitt}, B.~{DeWitt},  57--214

\bibitem[{Carter}(2010)]{Carter10}
{Carter} B., 2010, Gen Relativ Gravit, 42, 653

\bibitem[{Crinquand} et~al.(2020){Crinquand}, {Cerutti}, {Philippov}, {Parfrey}
  \& {Dubus}]{Crinquand20}
{Crinquand} B., {Cerutti} B., {Philippov} A., {Parfrey} K., {Dubus} G., 2020,
  \prl, 124, 14, 145101

\bibitem[{King} et~al.(1975){King}, {Lasota} \& {Kundt}]{King75}
{King} A.~R., {Lasota} J.~P., {Kundt} W., 1975, \prd, 12, 10, 3037

\bibitem[{King} \& {Pringle}(2021)]{KP21}
{King} A.~R., {Pringle} J.~E., 2021, arXiv e-prints,  arXiv:2107.12384

\bibitem[{Komissarov}(2004)]{ssk-ebh}
{Komissarov} S.~S., 2004, \mnras, 350, 427

\bibitem[{Komissarov} \& {McKinney}(2007)]{ssk-meissner}
{Komissarov} S.~S., {McKinney} J.~C., 2007, \mnras, 377, L49

\bibitem[{Levin} et~al.(2018){Levin}, {D'Orazio} \& {Garcia-Saenz}]{Levin18}
{Levin} J., {D'Orazio} D.~J., {Garcia-Saenz} S., 2018, \prd, 98, 12, 123002

\bibitem[{Nathanail} \& {Contopoulos}(2014)]{NC14}
{Nathanail} A., {Contopoulos} I., 2014, \apj, 788, 2, 186

\bibitem[{Pan}(2018)]{Pan18}
{Pan} Z., 2018, \prd, 98, 4, 043023

\bibitem[{Pan} \& {Yu}(2016)]{PY16}
{Pan} Z., {Yu} C., 2016, \apj, 816, 2, 77

\bibitem[{Parfrey} et~al.(2019){Parfrey}, {Philippov} \& {Cerutti}]{Parfrey19}
{Parfrey} K., {Philippov} A., {Cerutti} B., 2019, \prl, 122, 3, 035101

\bibitem[{Thorne} et~al.(1986){Thorne}, {Price} \& {MacDonald}]{TPM86}
{Thorne} K.~S., {Price} R.~H., {MacDonald} D.~A., 1986, {Black holes: The
  membrane paradigm}, New Haven: Yale University Press

\bibitem[{Wald}(1974)]{Wald74}
{Wald} R.~M., 1974, \prd, 10, 6, 1680

\end{thebibliography}


\end{document}